\documentclass[aps,prb,twocolumn]{revtex4}
\usepackage{epsfig}

\begin{document}
\headsep 2.5cm
\title{Real space Dynamical Super Cell Approximation for interacting  disordered systems.}

\author{Rostam Moradian}
\email{rmoradian@razi.ac.ir}
\affiliation{$^{1}$Physics Department, Faculty of Science, Razi
University, Kermanshah, Iran\\
$^{2}$Computational Physical Science Research Laboratory, Department of Nano-Science, Institute for Studies in Theoretical Physics and Mathematics (IPM)
            ,P.O.Box 19395-1795, Tehran, Iran}

\date{\today}

\begin{abstract}
Effective medium super-cell approximation method which is introduced for disordered systems is extended to a general case of interacting disordered systems. We found that the dynamical cluster approximation (DCA) and also the non local coherent potential approximation (NLCPA) are two simple case of this technique. Whole equations of this formalism derived by using the effective medium theory in real space.

\end{abstract}
\pacs{Pacs. 74.62.Dh, 74.70.Ad, 74.40.+k}

\maketitle
%\section{ Introduction }
Theoretical understanding of strongly correlated systems such as high temperature superconductors, heavy fermions and magnetism require appropriate techniques to obtain their physical properties. Recently, by the many body theory techniques, it has been shown that both alloys and strongly correlated systems in the infinite limit dimensions are mapped to a single site which embedded in an effective medium\cite{Metzner:89,Muller:89,Vlaming:92, moradian:00}. This can be described by restriction on locality of self energy, that is, $\Sigma(i,j;\imath\omega_{n})=\Sigma(\imath\omega_{n})\delta_{ij}$. This single site technique for strongly correlated systems and alloys systems called, {\em  Dynamical Mean Field Approximation} (DMFA)\cite{Georges:96} and {\em Coherent Potential Approximation} (CPA)\cite{Soven:67} respectively. The single site nature of the infinite dimensional limit is imply that the inter sites  correlations and also inter-site multiple scattering are  negligible. But in the ordinary dimensions such as one, two and three, both inter sites correlation and inter-site multiple scattering have a significant contribution on the self energy, therefore not only the self energy is not local but also it is very sensitive with respect to the dimension. Recently by coarse graining of the self energy in the k-space, the non-local correction to the DMFA has been considered, this method called the {\em Dynamical Cluster Approximation} (DCA)\cite{Hetler:98,Jarrell1:01}. But what is the correspondence of this k-space coarse graining in the real space was unknown. Also for a disordered system we have introduced a new real space technique that called effective medium super-cell approximation (EMACA)\cite{Moradian1:03}. It has been shown that this method leads to the coarse graining of self energy and also to the average Green function in the k-space. Also EMSCA recovers CPA, where number of sites in the super-cell is one ($N_{c}=1$), and is exact for the case of $N_{c}\rightarrow\infty$. Now we should ask this question: dose it exist a unique EMSCA, such that treats both interacting and randomness together?. We answer this question by introduce a general version of the EMSCA, where in a especial case of only interacting, it is reduce to the DCA\cite{Hetler:98} and for disordered systems leads to the old EMSCA\cite{Moradian1:03, Moradian1:04}.

In this paper the real space {\em effective medium super-cell approximation} (EMSCA)\cite{Moradian1:03, Moradian1:04} is generalized and extended to treat a random interacting system. We show that neglecting of the, interaction between electrons in different super-cells, multiple scattering by sites in different super-cells and also deviations of the hopping integrals, $\delta t^{\sigma\sigma^{'}}_{ij}$, naturally leads to the super-cell  periodicity of the self energy, $\Sigma(i,j;\imath\omega_{n})$, with respect to the super-cell translation vectors, ${\bf r}_{ij}$. This provide us the DCA\cite{Hetler:98} coarse graining of the self energy in k-space. We are obtain a closed set of equations to do calculations in a general EMSCA. Also whole relevant DCA equations are derived by the EMSCA method.

We start our investigation on a general tight binding model for an interacting alloy system, which is given by,       
\begin{eqnarray}
H&=&-\sum_{ij\sigma\sigma^{'}}t^{\sigma\sigma^{'}}_{ij}c^{\dagger}_{i\sigma}c_{j\sigma^{'}}
\nonumber\\&+&\sum_{i\sigma} (\varepsilon_{i}-\mu)
 \hat{n}_{i\sigma}+\sum_{ij\sigma\sigma^{'}} U^{\sigma\sigma^{'}}_{ij}\hat{n}_{i\sigma}\hat{n}_{j\sigma^{'}},
\label{eq:Hamiltonian}
\end{eqnarray}
where $c^{\dagger}_{i\sigma}$ ($c_{i\sigma}$) is the creation (annhilation) oprator of an electron with spin $\sigma$ on lattice site $i$ and $\hat{n}_{i\sigma}=c^{\dagger}_{i\sigma}c_{i\sigma}$ is the number oprator. $t^{\sigma\sigma^{'}}_{ij}$ are the random hopping integrals between $i$ and $j$ lattice sites with spin $\sigma$ and $\sigma^{'}$ respectively. $\mu$ is the chemical potential and $\varepsilon_{i}$ is the random on-site energy, where takes  $-\delta/2$ with probability  $1-c$ for the host sites and $\delta/2$ with probability $c$ for impurity sites. $U^{\sigma\sigma^{'}}_{ij}$ is a positive or negative interaction potential between electrons on the lattice site $i$ and  $j$.      

The equation of motion for electrons corresponding to the above Hamiltonian, Eq.\ref{eq:Hamiltonian}, is given by, 
\begin{eqnarray}
\sum_{l\sigma^{''}} \left(
       \begin{array}{c}
(\frac{\partial}{\partial\tau}-\varepsilon_{i}+\mu)\delta_{il}\delta_{\sigma\sigma^{''}}-t^{\sigma\sigma^{''}}_{il}\end{array}\right){ G}^{\sigma^{''}\sigma^{'}}(l\tau,j\tau^{'})
&+&\nonumber\\\sum_{l\sigma^{''}} 
U^{\sigma\sigma^{''}}_{lj}{ G}^{\sigma^{''}\sigma^{'}}_{2}(l\tau,l\tau,l\tau^{+},j\tau^{'})=\delta(\tau-\tau^{'})\delta_{ij}\delta_{\sigma\sigma^{'}}
\label{eq:equation of motion}
\end{eqnarray}
where $\tau$ and $\tau^{'}$ are imaginary time, $G^{\sigma\sigma^{'}}(i\tau,j\tau^{'})$ is the random interacting single particle Green function and $G^{\sigma\sigma^{'}}_{2}(i\tau,i\tau,i\tau^{+},j\tau^{'})$ is the two particle Green function. The random hopping integrals, $t^{\sigma\sigma^{'}}_{ij}$, can be defined in terms of clean system hoppings, $t^{0\sigma\sigma^{'}}_{ij}$, and  the hopping integral deviations, $\delta t^{\sigma\sigma^{'}}_{ij}$, in a such way that the hopping randomness is includes just in the $\delta t^{\sigma\sigma^{'}}_{ij}$, where
\begin{equation}     
t^{\sigma\sigma^{'}}_{ij}=t^{0\sigma\sigma^{'}}_{ij}+\delta t^{\sigma\sigma^{'}}_{ij}.
\label{eq:random t}
\end{equation} 
The Dyson equation corresponding to Eq.\ref{eq:equation of motion} for the averaged Green function, ${\bar G}^{\sigma\sigma^{'}}(l\tau^{''},j\tau^{'})$, is\cite{Gross:91},
\begin{eqnarray} 
\sum_{l\sigma^{''}} \left(
       \begin{array}{c}
(\frac{\partial}{\partial\tau}-\varepsilon_{i}+\mu)\delta_{il}\delta_{\sigma\sigma^{''}}-t^{0\sigma\sigma^{''}}_{il}\end{array}\right){\bar G}^{\sigma^{''}\sigma^{'}}(l\tau,j\tau^{'})
&+&\nonumber\\
\sum_{l\sigma^{''}}\int d\tau^{''}\Sigma^{\sigma\sigma^{''}}(i\tau,l\tau^{''}){\bar G}^{\sigma^{''}\sigma^{'}}(l\tau^{''},j\tau^{'})&=&\nonumber\\\delta(\tau-\tau^{'})\delta_{ij}\delta_{\sigma\sigma^{'}}
\label{eq:imaginary time equation of motion}
\end{eqnarray}
where the self energy, $\Sigma^{\sigma\sigma^{'}}(i\tau,l\tau^{'})$, is defined by,
\begin{eqnarray}
\sum_{l\sigma^{''}} \left(
       \begin{array}{c}
\langle(\varepsilon_{i}\delta_{il}\delta_{\sigma\sigma^{''}}+\delta t^{\sigma\sigma^{''}}_{il}\end{array}\right){ G}^{\sigma^{''}\sigma^{'}}(l\tau,j\tau^{'})
&+&\nonumber\\\sum_{l\sigma^{''}} 
U^{\sigma\sigma^{''}}_{lj}{ G}^{\sigma^{''}\sigma^{'}}_{2}(l\tau,l\tau,l\tau^{+},j\tau^{'})\rangle&=&\nonumber\\\sum_{l\sigma^{''}}\int d\tau^{''}\Sigma^{\sigma\sigma^{''}}(i\tau,l\tau^{''}){\bar G}^{\sigma^{''}\sigma^{'}}(l\tau^{''},j\tau^{'}).
\label{eq:self energy definition}
\end{eqnarray}
The imaginary time Fourier transform of Eq.\ref{eq:imaginary time equation of motion} leads to, 
\begin{eqnarray}
\sum_{l\sigma^{''}} \left(
       \begin{array}{c}
(\imath\omega_{n}+\mu)\delta_{il}\delta_{\sigma\sigma^{''}}-t^{0\sigma\sigma^{''}}_{il}-\Sigma^{\sigma\sigma^{''}}(i,l;\imath\omega_{n})\end{array}\right)\times\nonumber\\{\bar  G}^{\sigma^{''}\sigma^{'}}(l,j;\imath\omega_{n})=\delta_{ij}\delta_{\sigma\sigma^{'}},
\label{eq:Dyson1 equation}
\end{eqnarray}
where $\omega_{n}=\frac{1}{\beta}(2n+1)\pi$, are the Matsubura frequencies. Also the space Fourier transform of the Eq.\ref{eq:Dyson1 equation} is given by,
\begin{eqnarray}
\sum_{\sigma^{''}} \left(
       \begin{array}{c}
(\imath\omega_{n}+\mu)\delta_{\sigma\sigma^{''}}-\epsilon^{\sigma\sigma^{''}}_{\bf k}-\Sigma^{\sigma\sigma^{''}}({\bf k};\imath\omega_{n})\end{array}\right)\times\nonumber\\{\bar  G}^{\sigma^{''}\sigma^{'}}({\bf k};\imath\omega_{n})=\delta_{\sigma\sigma^{'}},
\label{eq:k-space average single particle green function}
\end{eqnarray}
where $\Sigma^{\sigma\sigma^{''}}({\bf k};\imath\omega_{n})$ is,
\begin{equation}
\Sigma^{\sigma\sigma^{'}}({\bf k};\imath\omega_{n})=\frac{1}{N}\sum_{i,j}\Sigma^{\sigma\sigma^{'}}(i,j;\imath\omega_{n}) e^{\imath{\bf k}.{\bf r}_{ij}}
\label{eq:self energy fourier transform}
\end{equation}
and also the clean system band structure, $\epsilon^{\sigma\sigma^{''}}_{\bf k}$ is defined by, 
 \begin{equation}
\epsilon^{\sigma\sigma^{''}}_{\bf k}=\frac{1}{N}\sum_{i,j}t^{0\sigma\sigma^{''}}_{ij} e^{\imath{\bf k}.{\bf r}_{ij}}.
\end{equation}
 The matrix form of Eq.\ref{eq:k-space average single particle green function} in the spinor space can be written as,
 \begin{equation}
{\bf \bar G}({\bf k};\imath\omega_{n})=\left(\begin{array}{c}
(\imath\omega_{n}+\mu){\bf I}-$$\mbox{\boldmath$\epsilon$}$$_{\bf k}-{\bf\Sigma}({\bf k};\imath\omega_{n}) 
\end{array}\right)^{-1} .
\label{eq:matix k-space average single particle green function}
\end{equation}
where the band structure matrix, $ $$\mbox{\boldmath$\epsilon$}_{\bf k}$$ $, is
\begin{equation}
\mbox{\boldmath$\epsilon$}_{\bf k}=\left(
       \begin{array}{cc}
\epsilon^{\uparrow\uparrow}_{\bf k} & \epsilon^{\uparrow\downarrow}_{\bf k} \\
 \epsilon^{\downarrow\uparrow}_{\bf k} & \epsilon^{\downarrow\downarrow}_{\bf k}\end{array}\right),
\label{eq: hopping integral matrix}
\end{equation}
and ${\bf\Sigma}({\bf k};\imath\omega_{n})$ is the self energy matrix, 
\begin{equation}
{\bf\Sigma}({\bf k};\imath\omega_{n})=\left(
       \begin{array}{cc}
\Sigma^{\uparrow\uparrow}({\bf k};\imath\omega_{n})& \Sigma^{\uparrow\downarrow}({\bf k};\imath\omega_{n}) \\
 \Sigma^{\downarrow\uparrow}({\bf k};\imath\omega_{n}) & \Sigma^{\downarrow\downarrow}({\bf k};\imath\omega_{n})\end{array}\right),
\label{eq: self energy matrix}
\end{equation}
 and the average Green function matrix, ${\bf G}({\bf k};\imath\omega_{n})$, is defined by, 
\begin{equation}
{\bf G}({\bf k};\imath\omega_{n})=\left(
       \begin{array}{cc}
G^{\uparrow\uparrow}({\bf k};\imath\omega_{n})& G^{\uparrow\downarrow}({\bf k};\imath\omega_{n}) \\
 G^{\downarrow\uparrow}({\bf k};\imath\omega_{n}) & G^{\downarrow\downarrow}({\bf k};\imath\omega_{n})\end{array}\right),
\label{eq: average Green matrix}
\end{equation}
also, ${\bf I}$ is a $2\times 2$ unitary matrix.
Hence the clean non-interacting Green function is given by,
 \begin{equation}
{\bf G}^{0}({\bf k};\imath\omega_{n})=\left(\begin{array}{c}
(\imath\omega_{n}+\mu){\bf I}-$$\mbox{\boldmath$\epsilon$}$$_{\bf k}  
\end{array}\right)^{-1} .
\label{eq:clean noninteracting single particl green function}
\end{equation}
 So, the real space correspondence of Eq.\ref{eq:k-space average single particle green function} can be written as, 
\begin{eqnarray}
 \bar{\bf G}(i,j;&\imath\omega_{n}&)={\bf G}^{0}(i,j;\imath\omega_{n})+\nonumber\\&\sum_{ll^{'}}&{\bf G}^{0}(i,l;\imath\omega_{n})
{\bf\Sigma}(l,l^{'};\imath\omega_{n})\bar{\bf G}(l^{'},j;\imath\omega_{n}),
\label{eq:real space Dyson equation}
\end{eqnarray}
where the real space clean non-interacting Green function, ${\bf G}^{0}(i,j;\omega_{n})$, is,
\begin{equation}
{\bf G}^{0}(i,j;\imath\omega_{n})=\frac{1}{N}\sum_{\bf k}e^{-\imath{\bf k}.{\bf r}_{ij}}{\bf G}^{0}({\bf k};\imath\omega_{n}),
\label{eq:real space clean}
\end{equation}
and the average single particle Green function is defined by,
 \begin{equation}
 \bar{\bf G}(i,j;\imath\omega_{n})=\frac{1}{N}\sum_{\bf k}e^{-\imath{\bf k}.{\bf r}_{ij}}{\bf \bar G}({\bf k};\imath\omega_{n}).
\label{eq:real space average green function}
\end{equation}
 
 Eq.\ref{eq:real space Dyson equation} can not be solved exactly. We extend  the {\em effective medium super-cell approximation} (EMSCA) which is recently introduced by us for a disordered system\cite{Moradian1:03}, to the case of interacting disordered systems. The EMSCA for such system based on three assumptions: first, {\em neglecting interaction between electrons on different super-cells},
\begin{equation}
{U}^{\sigma\sigma^{'}}_{ij} =0,\;\; if\;\; i\; and \;j\; \notin \;same\; super- cell,
\label{eq:interc poten}
\end{equation}
 second, {\em neglecting hopping integral deviations}, $\delta t^{\sigma\sigma^{'}}_{ij}$, when $i$ and $j$ are in the different super-cells,
\begin{equation}
 \delta t^{\sigma\sigma^{'}}_{ij} =0,\;\; if\;\; i\; and \;j\; \notin \;same\; super-cell.
\label{eq:different sites self energy}
\end{equation}
 and finally, {\em neglecting multiple impurity scattering  and also correlations between different super-cells}. These conditions imply that no correlation between super-cells, hence we have,
\begin{equation}
{\bf\Sigma}_{sc}(i,j;\imath\omega_{n}) =0,\;\; if\;\; i\; and \;j\; \notin \;same\; super-cell.
\label{eq:different sites self energy}
\end{equation}
 This means that the self energies in each super-cell are independent of other super-cells and they are periodic with respect to the super-cell translation vectors, ${\bf r}_{Nc}$,
 \begin{equation}
{\bf\Sigma}_{sc}({\bf r}_{IJ}+{\bf r}_{Nc};\imath\omega_{n})={\bf\Sigma}_{sc}({\bf r}_{IJ};\imath\omega_{n}), 
\label{eq:self periodicity}
\end{equation}
where $I$ and $J$ refer to sites in a same super-cell. Also Eq.\ref{eq:interc poten} leads to the super-cell periodicity for the interaction potential matrix, ${\bf U}_{sc}(I,J)$, where
\begin{equation}
{\bf U}_{sc}(I,J)=\left(
       \begin{array}{cc}
U^{\uparrow\uparrow}_{sc}(I,J)& U^{\uparrow\downarrow}_{sc}(I,J) \\
 U^{\downarrow\uparrow}_{sc}(I,J) & U^{\downarrow\downarrow}_{sc}(I,J) \end{array}\right),
\label{eq: potential matrix}
\end{equation}
 which is,
 \begin{equation}
{\bf U}_{sc}({\bf r}_{IJ}+{\bf r}_{Nc})={\bf U}_{sc}({\bf r}_{IJ}). 
\label{eq:pot periodicity}
\end{equation}
 The Fourier transformation of Eqs.\ref{eq:self periodicity} and \ref{eq:pot periodicity} imply that\cite{Moradian1:03},
 \begin{equation}
e^{-\imath{\bf k}.{\bf r}_{Nc}}=1.  
\label{eq:self-pot periodicity}
\end{equation}
Thus the wave vectors of the self energy, ${\bf\Sigma}({\bf k};\imath\omega_{n})$, and also of the interaction potential, ${\bf U}({\bf k})$ are restricts to the ${\bf K}_{n}$ which are given by,
\begin{equation}
{\bf K}_{n}=\sum_{i=1}^{3} \frac{l_{i}}{Nc_{i}} {\bf b_{i}},
\label{eq:supercell wave vector}
\end{equation}
where ${\bf b_{i}}$ are the reciprocal lattice primitive vectors, $Nc_{i}$ are the number of the lattice sites in a super-cell at ${\bf a_{i}}$ direction (${\bf a_{i}}$ are the lattice primitive vectors) and $l_{i}$ is a integer number. Therefore by inserting Eqs.\ref{eq:supercell wave vector} and \ref{eq:different sites self energy} in to Eq.\ref{eq:self energy fourier transform} we found that,
\begin{equation}
{\bf\Sigma}_{sc}({\bf K}_{n};\imath\omega_{n})=\frac{1}{Nc}\sum_{I,J}e^{\imath{\bf K}_{n}.{\bf r}_{IJ}}{\bf\Sigma}_{sc}(I,J;\imath\omega_{n}). 
\label{eq:K-space self energy supercell}
\end{equation}
Also by applying the EMSCA conditions, Eqs.\ref{eq:interc poten} and \ref{eq:self-pot periodicity}, to the following exact relation,
\begin{equation}
{\bf U}({\bf k})=\frac{1}{N}\sum_{i,j}{\bf U}(i,j) e^{\imath{\bf k}.{\bf r}_{ij}}, 
\label{eq:exact k-space self energy}
\end{equation}
we found a similar coarse graining for the interaction potential, 
\begin{equation}
{\bf U}({\bf K}_{n})=\frac{1}{N_{c}}\sum_{I,J}{\bf U}_{sc}(I,J) e^{\imath{\bf K}_{n}.{\bf r}_{IJ}}. 
\label{eq:coarse grained self energy}
\end{equation}
The inverse Fourier transformation of the self energy, ${\bf\Sigma}({\bf K}_{n};\imath\omega_{n})$, to the real super-cell is,
\begin{equation}
{\bf\Sigma}_{sc}(I,J;\imath\omega_{n})=\frac{1}{Nc}\sum_{{\bf K}_{n}}e^{-\imath{\bf K}_{n}.{\bf r}_{IJ}}{\bf\Sigma}({\bf K}_{n};\imath\omega_{n}). 
\label{eq:real space self energy supercell}
\end{equation}
and also for ${\bf U}({\bf K}_{n})$ is,
\begin{equation}
{\bf U}_{sc}(I,J)=\frac{1}{Nc}\sum_{{\bf K}_{n}}{\bf U}({\bf K}_{n}) e^{-\imath{\bf K}_{n}.{\bf r}_{IJ}}, 
\label{eq:real space potential supercell}
\end{equation}
where the orthogonality condition in a super-cell is given by,       
\begin{equation}
\frac{1}{N_{c}}\sum_{{\bf K}_{n}}e^{-\imath{\bf K}_{n}.{\bf r}_{IJ}} =\delta_{IJ}.
\label{eq:supercell orthogonality}
\end{equation}
Now by inserting Eq.\ref{eq:K-space self energy supercell} and \ref{eq:self-pot periodicity} in to  Eq.\ref{eq:real space average green function}, the super-cell average Green function, $\bar{\bf G}_{sc}(I,J;\imath\omega_{n})$, is given by\cite{Moradian1:03},
 \begin{equation}
\bar{\bf G}_{sc}(I,J;\imath\omega_{n})=\frac{1}{Nc}\sum_{{\bf K}_{n}}e^{\imath{\bf K}_{n}.{\bf r}_{ij}}
\bar{\bf G}({\bf K}_{n};\imath\omega_{n})
\label{eq:supercell average green function}
\end{equation}
where
 \begin{eqnarray}
\bar{\bf G}({\bf K}_{n};\imath\omega_{n})=\nonumber\\
\frac{N}{Nc}
\sum_{{\bf k}^{'}_{n}}\left(\begin{array}{c}
{\bf G}^{0}({\bf K}_{n}+{\bf k}^{'};\imath\omega_{n})^{-1}-\Sigma({\bf K}_{n};\imath\omega) 
\end{array}\right)^{-1} .
\label{eq:k-supercell average green function}
\end{eqnarray}
In order to obtain  $\bar{\bf G}_{sc}(I,J;\imath\omega_{n})$ from Eq.\ref{eq:supercell average green function}, we should have ${\bf\Sigma}_{sc}(I,J;\imath\omega_{n})$, thus we need to have another equations to complete the self consistency loop. These equations are obtained by applying the
% {\em Quantum Monte Carlo} (QMC) technique+
 EMSCA to the system partition function, as follow. The partition function of the system with the Hamiltonian Eq.\ref{eq:Hamiltonian} is given by,
\begin{equation}
Z=\langle \; Tr\; e^{-\beta {\hat H}}\;\rangle_{r},
\label{eq:patrition function}
\end{equation}
where $\langle\;\;\rangle_{r}$ denotes the configurational average over random energies, $\varepsilon_{i}$.
 The partition function, Eq.\ref{eq:patrition function}, can be rewrite as\cite{Imada:98} 
\begin{equation}
Z=\langle \int {\mathcal{ D}}{\bar \Psi} {\mathcal{  D}}\Psi  e^{-S} \;\rangle_{r},
\label{eq:1patrition function}
\end{equation}
where the action $S$ is,
\begin{equation}
S=\sum_{ij\sigma\sigma^{'}}\int_{0}^{\beta} d\tau{\bar \psi}_{i\sigma}(\delta_{ij}\delta_{\sigma\sigma^{'}}(\frac{\partial}{\partial \tau}-\mu)+t^{0\sigma\sigma^{'}}_{ij}){ \psi}_{j\sigma}(\tau)+S_{r-i},
\label{eq:clean intraction action}
\end{equation}
and $S_{r-i}$ is,
\begin{eqnarray}
S_{r-i}&=&\sum_{ij\sigma\sigma^{'}}\int d\tau{\bar \psi}_{i\sigma}\psi_{i\sigma}U^{\sigma\sigma^{'}}_{ij}{\bar \psi}_{j\sigma^{'}}(\tau)\psi_{j\sigma^{'}}(\tau)\nonumber\\&+&\sum_{ij\sigma}\int d\tau{\bar \psi}_{i\sigma}(\tau)\varepsilon_{i}\delta_{ij}{ \psi}_{j\sigma}(\tau)
\\&+&\sum_{ij\sigma\sigma^{'}}\int d\tau{\bar \psi}_{i\sigma}(\tau)\delta t^{\sigma\sigma^{'}}_{ij}{ \psi}_{j\sigma^{'}}(\tau),
\label{eq:random interaction action}
\end{eqnarray}
in which $\mathcal{D}\Psi=\Pi_{i}d\psi_{i\sigma}d\psi_{i\sigma}$ and $\mathcal{D}{\bar\Psi}=\Pi_{i}d{\bar\psi}_{i\sigma}d{\bar\psi}_{i\sigma}$, where $d{\bar\psi}_{i\sigma}=\lim_{M\rightarrow\infty}\Pi^{M}_{m=1}d{\bar\psi}_{i\sigma}(\tau_{m})$, $d{\psi}_{i\sigma}=\lim_{M\rightarrow\infty}\Pi^{M}_{m=1}d{\psi}_{i\sigma}(\tau_{m})$.
The Eq.\ref{eq:clean intraction action} can be written as, 
\begin{equation}
S=\int_{0}^{\beta} d\tau d\tau^{'}\sum_{ij\sigma\sigma^{'}}{\bar \psi}_{i\sigma}(\tau)({{\bf G}^{0}}^{-1})_{ij\sigma\sigma^{'}}\psi_{j\sigma^{'}}(\tau^{'})+S_{r-i},
\label{eq:random action}
\end{equation}
where,  
\begin{eqnarray}
&(\delta_{ij}\delta_{\sigma\sigma^{'}}(&\frac{\partial}{\partial \tau}-\mu)+t^{0\sigma\sigma^{'}}_{ij}){ \psi}_{j\sigma^{'}}(\tau)=\int_{0}^{\beta}d\tau^{'}\times\nonumber\\& \frac{1}{\beta}&\sum_{\omega_{n}}\left(\delta_{ij}\delta_{\sigma\sigma^{'}}(\imath \omega_{n}-\mu)+t^{0\sigma\sigma^{'}}_{ij}\right)e^{\imath\omega_{n}(\tau-\tau^{'})}{ \psi}_{j\sigma}(\tau^{'})\nonumber\\
&=&\int_{0}^{\beta}d\tau^{'}({{\bf G}^{0}}^{-1})_{ij\sigma\sigma^{'}}(\tau-\tau^{'}){ \psi}_{j\sigma^{'}}(\tau^{'}),
\label{eq:clean non intraction Green function}
\end{eqnarray}
and the clean non-interacting Green function matrix, ${\bf G}^{0}(\imath\omega_{n})$, is defined by
\begin{eqnarray}
({{\bf G}^{0}}^{-1})_{ij\sigma\sigma^{'}}(\tau-\tau^{'})=
\frac{1}{\beta}\sum_{\omega_{n}}\left(\delta_{ij}\delta_{\sigma\sigma^{'}}(\imath \omega_{n}-\mu)+t^{0\sigma\sigma^{'}}_{ij}\right)\nonumber\\\times e^{\imath\omega_{n}(\tau-\tau^{'})}.\;\;
\label{eq:inverce clean green function}
\end{eqnarray}
By hint of Eq.\ref{eq:real space Dyson equation} the clean Green function, ${\bf G}^{0}$, can be express in terms of the average Green function, $\bar {\bf G}$ , and self energy, ${\bf \Sigma}$, as,
\begin{equation}
{{\bf G}^{0}}^{-1}={\bar{\bf G}}^{-1}+ {\bf \Sigma}.
\label{eq:clean interm of average Green function}
\end{equation}
 By inserting Eq.\ref{eq:clean interm of average Green function} in to Eq.\ref{eq:random action} we have, 
\begin{equation}
S=\int d\tau d\tau^{'}\sum_{ij\sigma\sigma^{'}}{\bar \psi}_{i\sigma}(\tau)\left({\bar{\bf G}}^{-1}\right)_{ij\sigma\sigma^{'}}\psi_{j\sigma^{'}}(\tau^{'})+S_{r-i-s},
\label{eq:2random action}
\end{equation}
where,
\begin{eqnarray}
S_{r-i-s}&=&\sum_{ij\sigma\sigma^{'}}\int d\tau{\bar \psi}_{i\sigma}\psi_{i\sigma}U^{\sigma\sigma^{'}}_{ij}{\bar \psi}_{j\sigma^{'}}(\tau)\psi_{j\sigma^{'}}(\tau)\nonumber\\&+&\sum_{ij\sigma\sigma^{'}}\int d\tau d\tau^{'}{\bar \psi}_{i\sigma}(\tau)(\varepsilon_{i}\delta_{ij}\delta_{\sigma\sigma^{'}}+\delta t^{\sigma\sigma^{'}}_{ij}){ \psi}_{j\sigma^{'}}(\tau)\nonumber\\&-&\sum_{ij\sigma\sigma^{'}}\int d\tau^{'} d\tau{\bar \psi}_{i\sigma}(\tau){\bf\Sigma}^{\sigma\sigma^{'}}_{ij}(\tau-\tau^{'}){ \psi}_{j\sigma}(\tau^{'}).\nonumber\\
\label{eq:self-random interaction action}
\end{eqnarray}

Now by applying the EMSCA\cite{Moradian1:03}, where is taking average over all super-cells except one super-cell, which is denoted by $\{I\}$, Eq.\ref{eq:1patrition function} converts to,
\begin{eqnarray}
&Z_{\tiny EMSCA}&= Z_{sc}\times\int \Pi_{i \notin \{I\},\sigma}\left(d{\bar\psi}_{i\sigma}d{\psi}_{i\sigma}\right)
\nonumber\\& \times& 
e^{-\int d\tau d\tau^{'}\sum_{\sigma\sigma^{'}}\sum_{ij \notin \{I,J\} }{\bar \psi}_{i\sigma}(\tau)\left({\bar{\bf G}}^{-1}\right)_{ij\sigma\sigma^{'}}\psi_{j\sigma^{'}}(\tau^{'})},\nonumber\\
\label{eq:2s-c-a patrition function}
\end{eqnarray} 
where the super-cell partition function, $Z_{sc}$, is given by,
\begin{equation}
Z_{sc}= \langle\int\Pi_{I=1}^{Nc,\sigma}\left(d{\bar\psi}_{I\sigma}d{\psi}_{I\sigma}\right)e^{-S^{sc}_{r-i}}  \;\rangle_{r-sc} 
\label{eq:sc patrition function}
\end{equation} 
and the super-cell action in the effective medium, $S^{sc}_{r-i}$, is,
\begin{eqnarray}
S^{sc}_{r-i}&=&\sum_{IJ\sigma\sigma^{'}}\int d\tau d\tau^{'}{\bar \psi}_{I\sigma}(\tau)\left({\mbox{\boldmath{$\mathcal {G}$}}}^{-1}\right)_{IJ\sigma\sigma^{'}}(\tau-\tau^{'}){ \psi}_{J\sigma^{'}}(\tau^{'})
\nonumber\\&-&
\sum_{IJ\sigma\sigma^{'}}\int d\tau{\bar \psi}_{I\sigma}\psi_{I\sigma}U^{\sigma\sigma^{'}}_{scIJ}{\bar \psi}_{J\sigma^{'}}(\tau)\psi_{J\sigma^{'}}(\tau)
\nonumber\\&+&
\sum_{IJ\sigma\sigma^{'}}\int d\tau{\bar \psi}_{I\sigma}(\tau)(\varepsilon_{I}\delta_{\sigma\sigma^{'}}\delta_{IJ}+\delta t^{\sigma\sigma^{'}}_{IJ}){ \psi}_{J\sigma^{'}}(\tau),
\label{eq:super cell random interaction action}
\end{eqnarray}
in which the super-cell cavity Green function matrix, $\mbox{\boldmath{$\mathcal {G}$}}$, is defined by,
\begin{equation}
{\mbox{\boldmath$\mathcal{ G}$}}^{-1}={\bar{\bf G}}^{-1}_{sc}-{\bf\Sigma}_{sc}.
\label{eq:super cell cavity}
\end{equation}
The matrix element of Eq.\ref{eq:super cell cavity} is given by the following Dysons like equation for the super-cell sites,
\begin{eqnarray}
&\bar{G}_{sc}&(I,J;\imath\omega_{n})=
{\mathcal {G}}(I,J;\imath\omega_{n})\nonumber\\&+&\sum_{L,L^{\prime }}{\mathcal{G}}(I,L;\imath\omega_{n})
\Sigma_{sc}(L,L^{\prime };\imath\omega_{n})\bar{G}_{sc}(L^{\prime},J;\imath\omega_{n}).\nonumber\\
\label{eq:coherent cluster Green function}
\end{eqnarray}
The second part of the right hand side of Eq.\ref{eq:2s-c-a patrition function} is the super-cell excluded effective medium partition function which is easily integrable, due to bilinearity of the Grassmann variables. But the first part is the partition function of the super-cell where is embedded in an effective medium environment, is not integrable directly due to four point Grassmann variable in its integrand.
Note that for the case of only disordered systems, Eq.\ref{eq:super cell random interaction action} leads to,
\begin{equation}
{{\bf G}^{imp}_{sc}}^{-1}(\tau-\tau^{'})=\mbox{\boldmath{$\mathcal{ G}$}}(\imath\omega_{n})^{-1}(\tau-\tau^{'})-\delta(\tau-\tau^{'})\mbox{ \boldmath$\varepsilon$}_{sc}
\label{eq:imp green}
\end{equation}
where $\mbox{\boldmath$\varepsilon$}_{sc}$ is the super-cell impurity matrix. The imaginary time Fourier transform of Eq.\ref{eq:imp green} imply that,
\begin{equation}
{{\bf G}^{imp}_{sc}(\imath\omega_{n})}^{-1}={\mbox{\boldmath$\mathcal{ G}$}(\imath\omega_{n})}^{-1}-{\mbox{\boldmath $\varepsilon$}}_{sc}
\label{eq:imp green function matrix}
\end{equation}
where the matrix element of Eq.\ref{eq:imp green function matrix} can be written as, 
\begin{eqnarray}
{ G}^{imp}_{sc}(I,J,\imath\omega_{n})&=&{\mathcal{ G}}(I,J,\imath\omega_{n})\nonumber\\&+&\sum_{L}{\mathcal{ G}}(I,L,\imath\omega_{n}) \varepsilon_{L}{ G}^{imp}_{sc}(L,J,\imath\omega_{n})\nonumber\\
\label{eq:matrix element of imp green}
\end{eqnarray}
which are derived previously by us.\cite{Moradian1:03}. Therefore, for a disordered system, Eqs.\ref{eq:supercell average green function},\ref{eq:coherent cluster Green function} and \ref{eq:matrix element of imp green} are construct a close set of equations that should be solved self consistently.

 In the general case, where is included both interaction and randomness, to calculate the super-cell partition function ,$Z_{sc}$, it is possible to use the Hirsch-Hubbard-Stratonovich transformation (HHST)\cite{Hirsch:83} to decouple the interaction term and map it to an auxiliary Ising filed. Although our method is general with respect to the interaction potential and randomness, but for simplicity and also to see the DCA derivation we concentrate our discussion on a repulsive on-site potential, $U^{\sigma\sigma^{'}}_{ij}=U\delta_{ij}\delta_{\sigma,-\sigma}$ and $\delta t^{\sigma\sigma^{'}}_{ij}=0$. The HHST procedure is as following, dividing the imaginary time interval, $[0,\beta]$, into $M$ subintervals, $\Delta\tau=\frac{\beta}{M}$, hence the imaginary time at $l$th slice is given by $\tau_{l}=l\frac{\beta}{M}$. Therefore the discertizing of imaginary times leads to $\int_{0}^{\beta} d\tau=\sum_{l}\Delta\tau$ \cite{Held:02}, thus,  
\begin{equation}
Z_{sc}= \langle\int\Pi_{\sigma}\Pi_{I=1}^{Nc}\left(d{\bar\psi}_{I\sigma}d{\psi}_{I\sigma}\sum_{\{s_{I=\pm 1}\}}\right)e^{-S^{sc}_{r-i}}  \;\rangle_{r-sc},
\label{eq:QMC sc patrition function}
\end{equation} 
where the super-cell action $S^{sc}_{r-i}$ is, 
\begin{eqnarray}
 S^{sc}_{r-i}=(\Delta\tau)^{2}\sum_{IJll^{'}\sigma}{\bar \psi}_{I\sigma}(\tau_{l})\times
\nonumber\\
 \left(({\mbox{\boldmath$\mathcal {G}$}}^{-1})_{IJll^{'}}+\delta_{IJ}(\frac{\lambda\sigma s_{Il}}{(\Delta\tau)^{2}}-\frac{\varepsilon_{I}}{\Delta\tau}+\frac{ U}{2\Delta\tau} )\delta_{ll^{'}+1}\right){ \psi}_{J\sigma}(\tau_{l^{'}}).
\nonumber\\
\label{eq:QMC sc action}
\end{eqnarray} 
We define the cluster impurity Green function ${\bf G}^{imp}_{sc}$ as
\begin{equation}
({{\bf G}^{imp}_{sc}}^{-1})_{IJll^{'}}=({\mbox{\boldmath$\mathcal {G}$}}^{-1})_{IJll^{'}}+\delta_{IJ}(\frac{\lambda\sigma s_{Il}}{(\Delta\tau)^{2}}-\frac{\varepsilon_{I}}{\Delta\tau}+\frac{ U}{2\Delta\tau} )\delta_{ll^{'}+1}.
\label{eq:impurity green function}
\end{equation} 
The matrix form of Eq.\ref{eq:impurity green function} can be written as, 
\begin{equation}
({{\bf G}^{imp}_{sc}}^{-1})={\mbox{\boldmath${\mathcal  G}$}}+{\bf T}(e^{{\bf V}_{\sigma}}-\bf 1),
\label{eq:impurity green function matrix}
\end{equation} 
where, ${\bf T}_{IJll^{'}}=\frac{\delta_{IJ}\delta_{ll^{'}}}{{(\Delta\tau)^{2}}}$ and matrix elements of ${\bf V}_{\sigma}$ are,
\begin{equation}
{\bf V}_{Il\sigma}={\lambda\sigma s_{Il}}-\varepsilon_{I}{\Delta\tau}+\frac{ U\Delta\tau}{2} .
\label{eq:HH Ising}
\end{equation} 
Eqs.\ref{eq:impurity green function matrix} and \ref{eq:impurity green function} for the case of just interacting systems are derived by the DCA method\cite{Jarrell1:01}. Details of QMC solving of Eq.\ref{eq:impurity green function matrix} can be find in elsewhere\cite{Jarrell1:01, maier:04}.

 The average of impurity Green function, ${\bf G}^{imp}_{sc}$, over Ising fields and also impurities configurations is the super-cell effective medium Green function
\begin{equation}
\langle{\bf G}^{imp}_{sc}(I,J;{\tau}_{l},{\tau}_{l^{'}})\rangle={\bar{\bf G}}_{sc}(I,J;{\tau}_{l},{\tau}_{l^{'}}).
\label{eq:average impurity green function}
\end{equation} 
The Fourier transform of Eq.\ref{eq:average impurity green function} is give by,
\begin{eqnarray}
{\bar{\bf G}}_{sc}(K_{n};\imath\omega_{n})=\frac{1}{N_{c}}\times\nonumber\\\sum_{ll^{'}}\sum_{IJ}e^{\imath{\bf K}_{n}.{\bf r}_{IJ}}e^{\imath\omega_{n}({\tau}_{l}-{\tau}_{l^{'}})}{\bar{\bf G}}_{sc}(I,J;{\tau}_{l},{\tau}_{l^{'}}).
\label{eq:fourier transform average impurity green function}
\end{eqnarray} 

Eqs.\ref{eq:k-supercell average green function}, \ref{eq:coherent cluster Green function} and \ref{eq:fourier transform average impurity green function} construct a close set of equations that should be solved self consistency.

The algorithm of numerical proses is as following,

1- A guess for the initial cluster self energy, ${\bf \Sigma}(K_{n};\imath\omega)$, usually zero.

2- From Eq.\ref{eq:k-supercell average green function} calculate the cluster Green function, ${\bf G}(K_{n};\imath\omega)$.

3- Calculate the cavity Green function ${\mbox{\boldmath$\mathcal {G}$}}({\bf K}_{n};\imath\omega)$ from Fourier transform of Eq.\ref{eq:coherent cluster Green function}, ${\mbox{\boldmath$\mathcal {G}$}}^{-1}({\bf K}_{n};\imath\omega_{n})={\bf G}^{-1}({\bf K}_{n};\imath\omega_{n})+{\bf\Sigma}({\bf K}_{n};\imath\omega_{n})$.

4- Calculate the Fourier transform of the cavity Green function,
\begin{equation}
{\mbox{\boldmath$\mathcal {G}$}}(I,J;\tau_{l}-\tau_{l^{'}})=\frac{1}{\beta}\sum_{n}\sum_{{\bf K}_{n}}{\mbox{\boldmath$\mathcal {G}$}}({\bf K}_{n};\imath\omega_{n})e^{\imath\omega_{n}}e^{-\imath{\bf K}_{n}.{\bf r}_{IJ} }
\end{equation}

5- calculate the new cluster Green function ${\bf\bar G}(I,J;\tau_{l}-\tau_{l^{'}})$ from Eq.\ref{eq:impurity green function matrix}, \ref{eq:fourier transform average impurity green function}.

6- Calculate the inverse Fourier transform of ${\bf\bar G}(I,J;\tau_{l}-\tau_{l^{'}})$

7- Calculate the new self energy ${\bf\Sigma}({\bf K}_{n};\imath\omega_{n})$ from 
\begin{equation}
{\bf\Sigma}({\bf K}_{n};\imath\omega_{n})={\mbox{\boldmath$\mathcal {G}$}}^{-1}({\bf K}_{n};\imath\omega_{n})-{\bf G}^{-1}({\bf K}_{n};\imath\omega_{n}).
\end{equation}

8- Go to step 2 and repeat whole proses until convergence.

In conclusion we are extended the EMSCA to the case of an interacting disordered system. Similar to the EMSCA we showed that the periodicity of self energy with respect to super-cell translation vector leads to the coarse graining of self energies and hence the average Green function in k-space. Then by applying the effective medium theory on the system partition function, we find two equations where are relates the super-cell impurity Green function, ${{{\bf G}^{imp}}_{sc}}(I,J;\imath\omega_{n})$, and super cell average Green function, ${\bar{\bf G}}_{sc}(I,J;\imath\omega_{n})$,  to the super-cell cavity Green function, $\mbox{\boldmath$\mathcal {G}$}(I,J;\imath\omega_{n})$. This completes the whole formalism of a new real space method for disordered interaction systems. In especial case of interacting system our formalism leads to real space derivation of DCA\cite{Hetler:98} while for a disoreded system is converts to the EMSCA\cite{Moradian1:03}. Now we are established that the DCA and also the NLCPA are two especial case of the, real space dynamical effective medium super-cell approximation (DEMSCA).
\acknowledgements I would like to thanks James F. Annett and B. L. Gyorffy for helpful discussions.

\end{document}